\documentclass[prb,aps,preprint,showpacs,preprintnumbers,floats,superscriptaddress]{revtex4}
\usepackage{amsmath,amssymb,amsfonts, amsthm}
\usepackage{subfigure}
\usepackage{graphicx}
\usepackage{epsfig}
\newlength{\upit}\upit=0.1truein

\newcommand{\ltappr}{{{\lower4pt\hbox{$<$} } \atop \widetilde{ \ \ \ }}}
\newlength{\bxwidth}\bxwidth=1.5 truein

\newcommand{\dg}{^{\dagger }}

\newcommand{\rarrow}{\rightarrow}

\newlength{\figwidth}
\newlength{\shift}
\newcommand \bea {\begin{eqnarray} }
\newcommand \eea {\end{eqnarray}}
\newcommand{\br}{{\bf{r}}}
\newcommand{\bk}{{\bf{k}}}

\newcommand{\bq}{{\bf{q}}}

\newcommand{\ca}{Ca_{2-x}Na_xCuO_2Cl_2}

\begin{document}
\title{Model for nodal quasiparticle scattering in a disordered vortex
lattice}
\date{\today}
\author{ Marianna Maltseva}
\email{maltseva@physics.rutgers.edu}
\author{P. Coleman}
\affiliation{
Center for Materials Theory,
Rutgers University, Piscataway, NJ 08855, U.S.A. } 

\begin{abstract}
Recent scanning tunneling experiments on $\ca$ by Hanaguri et al.\cite{Hanaguri} 
observe field-dependent quasiparticle interference
effects which are sensitive to the sign of the d-wave order parameter.
Their analysis of spatial fluctuations in the local density of states shows that
there is a selective enhancement of quasiparticle scattering events that
preserve the gap sign, and  a selective depression of the quasiparticle
scattering events that reverse the gap sign.
We introduce a model which accounts for 
this phenomenon as a consequence of 
vortex pinning to impurities. Each pinned vortex 
embeds several impurities in its core. The observations of recent
experiments 
can be  accounted  for 
by assuming that the scattering potentials of the impurities inside
the vortex cores acquire 
an additional resonant or Andreev scattering
component, both of which induce gap sign preserving scattering events.
\end{abstract}

%
\maketitle
%

\section{introduction}
Fundamental  studies of unconventional superconductors 
are currently hindered by the
scarcity of direct methods to determine the structure of the superconducting
order parameter. Apart from Josephson junction experiments, few
spectroscopic probes provide the valuable information about the phase of the order parameter.
In this work, we discuss how phase sensitive coherence effects can be studied
using scanning tunneling spectroscopy/microscopy (STS/STM).

The key idea is that the
evolution of the phase of the order parameter in momentum space can be
determined from the Fourier transformed  fluctuations in the tunneling
density of states. The sensitivity of these  fluctuations  to the scattering rates of superconducting
quasiparticles manifests itself through coherence factor effects. 
Quasiparticles in a superconductor are 
a coherent superposition of excitations of 
electrons and holes.
Coherence factors
characterize how the  scattering rate of a superconducting
quasiparticle
off a given scatterer differs from the  scattering rate of 
a bare electron off the same scatterer \cite{Tinkham}. 
Coherence factors  are determined by combinations of the Bogoliubov coefficients
$u_\bk$ and $v_\bk$,
which give proportions of the particle and hole components that constitute a
superconducting quasiparticle, 
\begin{eqnarray}\label{Bog}
c_{\bk\uparrow}=u_\bk a_{\bk\uparrow}+v_\bk a\dg_{-\bk\downarrow},\\
c_{\bk\downarrow}=-v_\bk a\dg_{\bk\downarrow}+u_\bk a_{-\bk\uparrow}.
\end{eqnarray}
 The momentum-dependent order parameter 
$\Delta_\bk=|\Delta_\bk|e^{i\phi(\bk)}$ has the same
sign as the Bogoliubov coefficient $v_\bk$, so that   studies of scattering rates 
of  quasiparticles with different momenta can delineate how the phase of the order
parameter $\phi(\bk)$ changes in momentum space.

In studies of unconventional superconductors with spatially varying order
parameter,
scanning tunneling spectroscopy provides 
 a spectroscopic probe  with a  real space resolution at the atomic level.
In the past, observation of  phase sensitive coherence effects with STM
 has been thwarted by the problem
of controlling the scatterers \cite{HoffmanThesis}.
An ingenious solution of this problem has been found in the application of a magnetic
field, which introduces vortices as controllable scatterers in a given system
\cite{Hanaguri}.

In this work, we develop a framework observation of
 coherence factor effects with Fourier Transform Scanning Tunneling Spectroscopy
(FT-STS). Using this framework, we   analyze  the  recent
observations of the coherence factor effects in a magnetic field to develop
a phenomenological model of quasiparticle scattering in a disordered vortex
array.

\section{Coherence factors in STM measurement}
Scanning tunneling spectroscopy, which involves tunneling of single
electrons between a scanning tip and a superconducting sample, 
offers an opportunity to examine 
how the spectrum of superconducting quasiparticles responds to disorder. 
We now discuss how we can extract phase-sensitive information from
STM data. 
\subsection{LDOS correlators $R^{even}$ and $R^{odd}$ have
well-defined coherence factors}

We describe
the electron field inside a superconductor  by a Balian-Werthammer
spinor \cite{BW}
\begin{gather*}
 \Psi(\br,\tau )=
\begin{pmatrix}
&\psi_{\uparrow} (\br,\tau)\\
&\psi_{\downarrow} ({\br},\tau )\\
&\psi_{\downarrow}^{\dagger} ({\br},\tau )\\
&-\psi_{\uparrow}^{\dagger} ({\br},\tau )
\label{Psi}
\end{pmatrix},
\end{gather*}
where $\br$ denotes real space coordinates and $\tau$ is imaginary time.
The Nambu Green's function is defined as the ordered average
\begin{equation}\label{matrixG}
\hat {G}_{\alpha \beta }({\br'},{\br};\tau )=
-\langle T_\tau \Psi_{\alpha }(\br',\tau ) \Psi_{\beta }^{\dagger}(\br,0)
\rangle .
\end{equation}
Tunneling measurements determine local density of states, which is
given by
\begin{equation}\label{rhoNambu}
\rho (\br,\omega) = 
\frac{1}{\pi} ~Im ~
{\rm Tr}~\frac{1+\tau_3}{2} \left[ G (\br,\br; \omega-i\delta )\right],
\end{equation}
where $G (\br',\br;z)$ is the analytic continuation 
$G (\br',\br;i\omega_{n})\rightarrow G (\br',\br;z)$ of the Matsubara
Green's function
\begin{equation}\label{GMats}
G (\br',\br;i\omega_{n})
 = \int_{0}^{\beta }G (\br',\br;\tau )e^{i\omega_{n}\tau }d\tau ,
 \end{equation}
with $\omega_n=(2n+1)\pi T$.
The appearance of the combination $\frac{1+\tau_3}{2}$ in (\ref{rhoNambu})
projects out the normal
component of the Nambu Green's function
\begin{equation}
Tr~\frac{1+\tau_3}{2} ~G (\br',\br;\tau )= -\sum_{\sigma}
\langle T_\tau \psi_{\sigma } (\br',\tau )\psi \dg_{\sigma } (\br,0)\rangle .
\end{equation}
The mixture of the unit and the $\tau_3$ matrices in this expression
prevents the local density of states from developing a well-defined
coherence factor.
We now show that the components of the local density of states
that have been symmetrized or antisymmetrized in the bias voltage
have a well-defined coherence factor.  The key result here is that
\begin{eqnarray}\label{odd-even}
\rho ({\br},\omega)
\pm 
\rho ({\br},-\omega) = \frac{1}{\pi}~ Im~ 
{\rm Tr}~ \left[\left\{\begin{array}{c}
1\cr
\tau_3
\end{array} \right\}
 G (\br,\br; \omega-i\delta )\right].
 \end{eqnarray}
In particular, this implies that the antisymmetrized density of states
has the same coherence factor as the charge density operator  $\tau_{3}$.

To show these results, we introduce the ``conjugation matrix''
$C= \sigma_{2}\tau_{2} $, whose action on the Nambu spinor is to
conjugate the fields,
\begin{equation}\label{conj}
C\Psi=  (\Psi \dg )^{T}\equiv \Psi^{*} ,
\end{equation}
effectively taking the Hermitian conjugate of each component of the
Nambu spinor.  This also implies that $\Psi \dg  C= \Psi^{T}$. Here 
$\tau_i$ are Pauli matrices acting in particle-hole space, for example,
\begin{gather*}
{\bf \tau_3}=
\begin{pmatrix}
&\underline{1}&0\\
&0&-\underline{1}
\label{tau3}
\end{pmatrix},
\end{gather*}
and $\sigma_i$ are Pauli matrices acting in spin space,
\begin{gather*}
{\bf \sigma_i}=
\begin{pmatrix}
&\underline{\sigma_i}&0\\
&0&\underline{\sigma_i}
\label{sigmai}
\end{pmatrix}.
\end{gather*}
Using (\ref{conj}), it follows that 
\begin{eqnarray}\label{l}
[C G(\br',\br;\tau )C]_{\alpha \beta } &=& - \langle T_\tau C\Psi (\br',\tau)\Psi\dg
(\br,0)C\rangle_{\alpha \beta } \cr
&=& - \langle T_\tau\Psi_{\alpha }^{*} (\br',\tau)\Psi_{\beta }^{T} (\br,0)\rangle\cr
&=&   \langle T_\tau\Psi_{\beta } (\br,0)\Psi_{\alpha }\dg  (\br',\tau)\rangle\cr
&=& - G_{\beta \alpha } (\br,\br',-\tau), 
\end{eqnarray}
or,  in the matrix notation, 
\begin{eqnarray}\label{CGC}
C G(\br,\br';\tau )C&=&  -G^{T} (\br',\br;-\tau),
\end{eqnarray}
which in turn implies for the Matsubara Green's function (\ref{GMats})
\begin{eqnarray}\label{CGCiomega}
C G(\br,\br';i \omega_n )C&=&  -G^{T} (\br',\br;-i \omega_n ).
\end{eqnarray}
For the advanced Green's function, which is related to the Matsubara Green's
function via analytic continuation, 
$G (\br,\br',i\omega_{n})\rightarrow G (\br,\br',z)$, we obtain
\begin{eqnarray}\label{CGCAdv}
C G(\br,\br'; \omega-i\delta )C&=&  -G^{T} (\br',\br;-\omega+i\delta ).
\end{eqnarray}
Using this result and the commutation relations  of  Pauli matrices, we obtain
\begin{eqnarray}
\rho ({\br},-\omega) = 
-\frac{1}{\pi}~ Im~ 
{\rm Tr}~ \frac{1+\tau_3}{2}~G (\br,\br; -\omega+i\delta )=\nonumber\\
 = \frac{1}{\pi}~ Im~ 
{\rm Tr}~ \frac{1+\tau_3}{2}~C~G^T (\br,\br; \omega-i\delta)~C=\nonumber\\
 = \frac{1}{\pi}~ Im~ 
{\rm Tr}~ \frac{1-\tau_3}{2}~G (\br,\br; \omega-i\delta).
\end{eqnarray}
Finally, we obtain
\begin{eqnarray}
\rho ({\br},\omega)
\pm 
\rho ({\br},-\omega) = \frac{1}{\pi}~ Im~ 
{\rm Tr}~ \left[\frac{1+\tau_3}{2}~
 G (\br,\br; \omega-i\delta )\pm  \frac{1-\tau_3}{2}~G (\br,\br;
 \omega-i\delta)\right]
 =\nonumber\\
 = \frac{1}{\pi}~ Im~ 
{\rm Tr}~ \left[\left\{\begin{array}{c}
1\cr
\tau_3
\end{array} \right\}
 G (\br,\br; \omega-i\delta )\right].
\end{eqnarray}

\subsection{Coherence factors in a BCS superconductor, T-matrix approximation}
Next, applying this result  to a BCS superconductor, we show that
in the t-matrix
approximation the coherence factors that 
arise in the conductance ratio
$Z(\bq,V)$ are given by the product of the coherence factors associated with
the charge operator and the scattering potential.

T-matrix approximation \cite{Balatsky, Hirschfeld-86}
allows to compute the  Green's function
 in the presence of multiple
scattering off impurities. In terms of the bare Green's function $G(\bk,\omega)$ and the
impurity t-matrix ${\hat t}(\bk,\bk')$, the full Green's function is given by 
\begin{eqnarray}\label{GT}
{\tilde G}(\bk,\bk',\omega)=G(\bk,\omega)+
G(\bk,\omega){\hat t}(\bk,\bk')G(\bk',\omega)=
G(\bk,\omega)+
\delta G(\bk,\bk',\omega).
\end{eqnarray}
Using this expression, we obtain for 
the Fourier transformed odd  fluctuations in the tunneling density of states
\begin{eqnarray}\label{deltarhooddT}
\delta \rho^{odd}(\bq,\omega)=\frac{1}{2\pi }
{\rm Im} \int_{\bk} {\rm Tr}\Bigl[\tau_3
\delta G_{\bk_{+},\bk_{-}} (\omega-i\delta )
\Bigr]=\nonumber\\=
\frac{1}{2\pi}
{\rm Im} \int_{\bk} {\rm Tr}\Bigl[\tau_3
G_{\bk_{-}} (\omega-i\delta )\hat{t}(\bq,\bk)
G_{\bk_{+}} (\omega-i\delta )
\Bigr].
\end{eqnarray}
The Fourier transformed even  fluctuations in the tunneling density of states
\begin{eqnarray}\label{deltarhoevenT}
\delta \rho^{even}(\bq,\omega)=\frac{1}{2\pi }
{\rm Im} \int_{\bk} {\rm Tr}\Bigl[
\delta G_{\bk_{+},\bk_{-}} (\omega-i\delta )
\Bigr]=\nonumber\\=
\frac{1}{2\pi}
{\rm Im} \int_{\bk} {\rm Tr}\Bigl[
G_{\bk_{-}} (\omega-i\delta )\hat{t}(\bq,\bk)
G_{\bk_{+}} (\omega-i\delta )
\Bigr].
\end{eqnarray}

For scattering off a single impurity with a scattering potential 
${\hat U}(\bk,\bk')$,
the t-matrix ${\hat t}(\bk,\bk')$ denotes the infinite sum
\begin{eqnarray}\label{t-matrix}
{\hat t}(\bk,\bk')={\hat U}(\bk,\bk')+\sum_{\bk''}{\hat U}(\bk,\bk'')
G(\bk'',\omega){\hat U}(\bk'',\bk')+...=\nonumber\\
={\hat U}(\bk,\bk')+\sum_{\bk''}{\hat U}(\bk,\bk'')
G(\bk'',\omega){\hat t}(\bk'',\bk').
\end{eqnarray}

Working in the Born approximation, which is equivalent to taking only the first
term in the series (\ref{t-matrix}), we derive the expressions for the coherence factors associated with 
some common scattering processes that arise  in the even and odd density-density
correlators
$R^{even}(\bq,V)$ and $R^{odd}(\bq,V)$ in a BCS superconductor (see Table 1).
We use the following expression for the BCS Green's function
 for an electron with a normal state dispersion
$\epsilon_{\bk }$ and a gap function $\Delta_{\bk }$: 
\begin{eqnarray}\label{GBCS}
G_{\bk } (\omega)= [\omega
-\epsilon_{\bk }\tau_{3}-\Delta_{\bk }\tau_{1}]^{-1},
\end{eqnarray}
$\hat{t}(\bq,\bk)$ is
the  scattering t-matrix of the impurity potential, and $\bk_{\pm }= \bk  \pm
\bq /2$.
If the scattering potential has the t-matrix given by 
$\hat{t}(\bq,\bk)=T_3(\bq)~\tau_3$,
corresponding to a weak scalar (charge) scatterer,
the change in the odd part of the Fourier transformed tunneling density of
states  becomes $\delta \rho^{odd}_{scalar}(\bq ,\omega)=
T_3(\bq)~\Lambda^{odd}_{scalar} (\bq ,\omega)$ with
\begin{eqnarray}\label{Lambdascodd}
\Lambda^{odd}_{scalar} (\bq ,\omega)= \frac{1}{2\pi }{\rm Im}\int_k~
\Bigl[
\frac{z^2+\epsilon_{\bk_+}\epsilon_{\bk_-}-\Delta_{\bk_+}\Delta_{\bk_-}}
{(z^2-E_{\bk_+}^2)
(z^2-E_{\bk_-}^2)}\Bigr]_{z=\omega-i\delta },
\end{eqnarray} 
where $E_{\bk}=[\epsilon_{\bk}^2+\Delta_{\bk}^2]^{\frac{1}{2}}$ is the
quasiparticle energy. 
Expressed in terms of the Bogoliubov coefficients $u_\bk$ and
$v_\bk$,  given by $u^2_k(v^2_k)=\frac{1}{2}(1\pm\epsilon_k/E_k)$,
the expression under the integral in (\ref{Lambdascodd})
 is proportional to $(u_+u_--v_+v_-)^2$.

Fluctuations in the even part of the Fourier transformed tunneling density of
states due to scattering off a scalar impurity are substantially smaller,
$R^{even}_{scalar}(\bq ,\omega)\ll R^{odd}_{scalar}(\bq
,\omega)$, where $R^{even(odd)}_{scalar}(\bq ,\omega)$ is defined by
(\ref{correven-odd}),
 $\delta \rho^{even}_{scalar}(\bq ,\omega)=
T_3(\bq)~\Lambda^{even}_{scalar} (\bq ,\omega)$ with
\begin{eqnarray}\label{Lambdasceven}
\Lambda^{even}_{scalar} (\bq ,\omega)= \frac{1}{2\pi }{\rm Im}\int_k~
\Bigl[
\frac{z(\epsilon_{\bk_+}+\epsilon_{\bk_-})}
{(z^2-E_{\bk_+}^2)
(z^2-E_{\bk_-}^2)}\Bigr]_{z=\omega-i\delta }.
\end{eqnarray} 
Expressed in terms of the Bogoliubov coefficients $u_\bk$ and
$v_\bk$,  
the expression under the integral in (\ref{Lambdasceven})
 is proportional to $(u_+u_-+v_+v_-)(u_+u_--v_+v_-)$, and is, therefore,
small for the nodal
 quasiparticles involved,
 $|\Lambda^{even}_{scalar}(\bq ,\omega)|\ll 
 |\Lambda^{odd}_{scalar}(\bq ,\omega)|$.
Thus, scattering off a weak scalar impurity 
contributes predominantly to odd-parity fluctuations in the density of
states,  $R^{odd}(\bq,V)$.

In a second example, consider scattering off a 
 pair-breaking ``Andreev" scatterer with the t-matrix given by 
$\hat{t}(\bq,\bk)=T_1(\bq,\bk)~\tau_1$.
Here
the change in the even and odd parts of the Fourier transformed tunneling
density of states are $\delta \rho_{A}^{even(odd)}(\bq ,\omega)=
\Lambda_{A}^{even(odd)} (\bq ,\omega)$ with
\begin{eqnarray}\label{LambdaAeven}
\Lambda_{A} ^{even}(\bq ,\omega)= \frac{1}{2\pi }{\rm Im}\int_k~T_1(\bq,\bk)~
\Bigl[
\frac{z(\Delta_{\bk_+}+\Delta_{\bk_-})}
{(z^2-E_{\bk_+}^2)
(z^2-E_{\bk_-}^2)}\Bigr]_{z=\omega-i\delta },\\
\label{LambdaAodd}
\Lambda_{A} ^{odd}(\bq ,\omega)= \frac{1}{2\pi }{\rm Im}\int_k~T_1(\bq,\bk)~
\Bigl[
\frac{\epsilon_{\bk_+}\Delta_{\bk_-}+\epsilon_{\bk_-}\Delta_{\bk_+}}
{(z^2-E_{\bk_+}^2)
(z^2-E_{\bk_-}^2)}\Bigr]_{z=\omega-i\delta }.
\end{eqnarray} In terms of the Bogoliubov coefficients $u_\bk$ and
$v_\bk$, 
the expressions in square brackets in $\Lambda_{A} ^{even}(\bq ,\omega)$
and $\Lambda_{A} ^{odd}(\bq ,\omega)$
are proportional to
$(u_+u_-+v_+v_-)(u_+v_-+v_+u_-)$ and $(u_+u_--v_+v_-)(u_+v_-+v_+u_-)$,
respectively. For the nodal
 quasiparticles involved, the latter expression is substantially smaller than
 the former,
 $|\Lambda^{odd}_{A}(\bq ,\omega)|\ll 
 |\Lambda^{even}_{A}(\bq ,\omega)|$.
Thus, scattering off an Andreev scatterer gives rise to mainly 
even parity fluctuations in the density of states, 
$R^{even}(\bq,V)$.

We summarize the coherence factors arising in $R^{even}(\bq,V)$ and $R^{odd}(\bq,V)$ for some common scatterers in Table 1. The dominant 
contribution for a particular type of scatterer is given in bold.

\begin{table}[h!b!p!]
\caption{Coherence factors C(q) in $R^{even}(\bq,V)$ and $R^{odd}(\bq,V)$ for some common scatterers. }
\begin{tabular}{llllll}
\hline
T-matrix~ ~& Scatterer & C(q) in $R^{even}(\bq,V)$  & C(q) in $R^{odd}(\bq,V)$
&Enhanced $q_i$ &Enhances "++"?\\
\hline
${\bf \tau_3}$~ & Weak Scalar~& $(uu'+vv')(uu'-vv')$ ~& ${\bf (uu'-vv')^2}$~&
2,3,6,7 ~&No\\
${\bf \sigma \cdot m}$~ & Weak Magnetic~& 0 ~& 0 ~& None ~& No\\
i sgn~$\omega$~${\bf \hat {1}}$~ & Resonant ~& ${\bf (uu'+vv')^2} $ ~&
$(uu'+vv')(uu'-vv')$~& 1,4,5 ~& Yes\\
${\bf \tau_1}$~ & Andreev ~&  $ {\bf (uu'+vv')(uv'+vu')}$~& 
$(uu'-vv')(uv'+vu')$ ~&1,4,5 ~& Yes\\
\hline
\end{tabular}
\label{table1}
\end{table}
\noindent

\noindent From Table 1, we see that 
the odd correlator $R^{odd}(\bq,V)$ is determined by a
product of coherence factors associated with the
 charge operator and the scattering potential, while the 
even correlator $R^{even}(\bq,V)$  is determined by a product of the coherence factors
associated with the unit operator and the scattering potential.



\subsection{Conductance ratio - measure of LDOS}
An STM experiment measures the differential tunneling conductance
$\frac{dI}{dV}(\br,V)$ at a location $\br$ and voltage $V$
\cite{NewReview}. In a simplified model of the tunneling,
\begin{equation}
\label{sigma}
\frac{dI}{dV}(\br,V)
\propto \int_{-eV}^0 d\omega [-f'(\omega-eV)]\int d\br_1d\br_2
M(\br_1,\br)M^*(\br_2,\br)A(\br_2,\br_1,\omega),
\end{equation}    
where $A(\br_2,\br_1,\omega)=\frac{1}{\pi}Im~ G(\br_2,\br_1,\omega-i\delta)$ is the single electron spectral function
and $f(\omega)$ is the Fermi function. 
Here $\br_1$, $\br_2$ and $\br$ are the two-dimensional coordinates of the
incoming and outgoing electrons, and the position of the tip, respectively.
$M(\br_1,\br)$
is the spatially dependent tunneling matrix element, which includes
contributions of the sample wave function around the tip. 

Assuming that the tunneling matrix element is local, we write
$M(\br_1,\br)=M(\br)\delta^{(2)}(\br_1-\br)$, where
$M(\br)$ is a smooth function of position $\br$. In the low-temperature
limit, when $T\rightarrow 0$, the derivative of the Fermi function is replaced by a
delta-function, $-f'(\omega-eV)=\delta(\omega-eV)$. With these simplifications,
 we obtain
\begin{equation}
\label{sigmaLoc}
\frac{dI}{dV}(r,V)\propto 
|M(\br)|^2\rho(\br,V),
\end{equation}    
where $\rho(\br,V)=A(\br,\br,V)$ is the single-particle density of states.
In the WKB approach the tunneling matrix element is given by
$|M(\br)|^2=e ^{-2\gamma(\br)}$ with $\gamma(\br)=
\int_0^{s(\br)} dx\sqrt{\frac{2m\psi(\br)}{\hbar^2}}
=\frac{s(\br)}{\hbar}\sqrt{2m\psi(\br)}$, where $s(\br)$ is the barrier width
(tip-sample separation), 
 $\psi(\br)$ is the barrier
height, which is a mixture of the work functions of the tip and the sample,
 $m$ is the electron mass \cite{NewReview,HoffmanThesis}. 
Thus, the tunneling conductance is a measure of the thermally smeared 
local  density of states (LDOS) of the sample at the position of the tip.
  
To filter out the spatial variations in the tunneling matrix elements
$M(\br)$, originating from local
variations in the barrier height $\phi$ and the tip-sample separation $s$,
the conductance ratio is taken:
\begin{equation}
\label{Z}
Z(r,V)=\frac{\frac{dI}{dV}(r,+V)}{\frac{dI}{dV}(r,-V)}=
\frac {\rho (r,+V)}{\rho (r,-V)}=
\frac { \rho_0(+V)+\delta\rho(r,+V) }{\rho_0(-V)+\delta\rho(r,-V) }.
\end{equation}    
For small fluctuations of the local density of
states, $\delta\rho(r,\pm V)\ll\rho_0(\pm V)$,
$Z(r,V) $ is given by a linear combination of positive and negative energy components
of the tunneling density of states,
\begin{equation}
\label{Zexpanded}
Z(r,V)\simeq Z_0 (V)~\Bigl[1+
\frac { \delta\rho(r,+V) }{\rho_0(+V) }-
\frac { \delta\rho(r,-V) }{\rho_0(-V) }\Bigl]
\end{equation} 
with $Z_0 (V)\equiv \frac{\rho_0(+V) }{\rho_0(-V) }$.
The Fourier transform of this quantity contains a single delta function term at
$\bq=0$ plus a diffuse background,
\begin{equation}
\label{Z(q,V)}
Z(\bq,V)=
Z_0 (V)(2\pi)^2 \delta^2(\bq)+Z_0 (V)~\Bigl[
\frac { \delta\rho(\bq,+V) }{\rho_0(+V) }-
\frac { \delta\rho(\bq,-V) }{\rho_0(-V) }\Bigr].
\end{equation}
Interference patterns produced by quasiparticle
scattering off impurities are observed in the diffuse background described by
the second term. 

Clearly, linear response theory is only valid when 
the fluctuations in the local
density of states are small compared with its average value,
$\overline{\delta\rho(r,\pm V)^{2}}\ll\rho_0(\pm V)^{2}$.
In the clean limit, this condition is satisfied at
finite and sufficiently large bias voltages $|V|>0$.
At zero bias voltage $V\rarrow
0$, however, the fluctuations in the local
density of states become larger than the vanishing density of states in the clean limit,
$|\delta\rho(r,\pm V)|>\rho_0(\pm V)$, and 
linear response theory can no longer be applied. 

At finite bias voltages, $|V|>0$, fluctuations in the conductance ratio $Z(\bq,V)$ are given by a sum
of two terms, even and odd in the bias voltage:
\begin{equation}
\label{Z-even-odd}
Z(\bq,V)|_{\bq\neq 0}=Z_0 (V)~\Bigl[
\delta\rho^{even}(\bq,V)(\frac {1}{\rho_0(+V) }-\frac {1}{\rho_0(-V) })+
\delta\rho^{odd}(\bq,V)(\frac {1}{\rho_0(+V) }+\frac {1}{\rho_0(-V) })\Bigr],
\end{equation}
where $\delta\rho^{even(odd)}(\bq,V)\equiv ( \delta\rho(\bq,+V) \pm
 \delta\rho(\bq,-V) )/2$.
 
Depending on the particle-hole symmetry properties of the
sample-averaged tunneling density of
states $\rho_0(V)$, one of these terms can dominate.
For example, if at the bias voltages used,
the sample-averaged tunneling density of
states $\rho_0(V)$ is approximately particle-hole symmetric,
$\rho_0(-V)\approx\rho_0(+V) =\rho_0(V)$, then $Z(\bq,V)$ is dominated by
the part of LDOS fluctuations that is odd in the bias voltage $V$,
\begin{equation}
\label{Zphsym}
Z(\bq,V)|_{\bq\neq 0}\simeq Z_0 (V)~\frac {2}{\rho_0(V) }
\delta\rho^{odd}(\bq,V).
\end{equation} 

In general, when we average over the impurity positions, the Fourier transformed 
fluctuations in the tunneling density of states, $\delta \rho
({\bq},V)$, vanish. However, the variance in the density of states fluctuations 
is non-zero and is given by the correlator
\begin{eqnarray}\label{corr}
R(\bq,V)&=&\overline{\delta \rho ({\bq},V)\delta \rho^*({-\bq},V)}.
\end{eqnarray} 
Defining 
\begin{eqnarray}  \label{correven-odd}
R^{even(odd)}(\bq,V)&=&\overline{\delta \rho^{even(odd)}({\bq},V)\delta 
\rho^{*even(odd)}
({-\bq},V)},
\end{eqnarray} we obtain that for $\bq\neq 0$
\begin{equation} \label{Zcorrodd}
|Z(\bq,V)|^2=\frac{4|Z_0(V)|^2}{\rho_0^2(V)}R^{odd}(\bq,V).
\end{equation} 
\subsection{Observation of coherence factor effects in QPI:
coherence factors and the octet model}

In high-Tc cuprates the quasiparticle interference (QPI) patterns,  observed in the
Fourier transformed tunneling conductance $dI(q,V)/dV\propto \rho(q,V)$, are dominated by a
small set of wavevectors $q_{1-7}$, connecting the ends of the banana-shaped
constant energy contours \cite{Hoffman, Howald,DHLee}.
This observation has been explained by the so-called ''octet'' model, which suggests
that the interference patterns are produced by elastic scattering off
random disorder between the  regions of the Brillouin zone with the largest
density of states, so that the scattering between the ends of the banana-shaped
constant energy contours, where the joint density of states is sharply peaked,
gives the dominant contribution to the quasiparticle interference patterns.

In essence, the octet model assumes that the fluctuations in the Fourier
transformed tunneling density of states are given by the following convolution: 
\[
\delta\rho(\bq,V)\propto\int_\bk\rho(\bk_+,\omega)\rho(\bk_-,\omega).
\]
While this assumption allows for a qualitative description, it is technically
incorrect
\cite{Pereg-Barnea-Franz,Scalapino}, for the correct expression for 
change in the density of states involves the imaginary part of a
product of Green's functions, rather than 
a product of the imaginary parts of the
Green's function, as written above. 
In this section, 
we show that the fluctuations in the conductance
ratio at wavevector $\bq $, given by $Z(\bq,V)$, are, 
nevertheless,  related to  the joint density of states via a Kramers-Kronig
transformation, 
so that the spectra of the conductance ratio $Z(\bq,V)$
can still be analyzed using the octet model.

As we have discussed, 
fluctuations in the density of states $\delta\rho(\bq,V)$ are determined by scattering off
impurity potentials and have the basic form (\ref{deltarhooddT}).
This quantity involves the imaginary part of a product of two Green's functions,
and as it stands, it is not proportional to the joint density of states. However, 
we can relate the two quantities by a Kramers-Kronig transformation, as we now
show.

We write the Green's function as
\begin{eqnarray}
G_\bk(E-i\delta)=\int\frac{d\omega}{\pi}\frac{1}{E-\omega-i\delta}
G''_\bk(\omega-i\delta),
\end{eqnarray}
where $G''_\bk(\omega-i\delta)=\frac{1}{2i}(G_\bk(\omega-i\delta)-
G_\bk(\omega+i\delta))$.
Substituting this form in (\ref{deltarhooddT}), we obtain
\begin{eqnarray}\label{deltarhooddTJ}
&\delta \rho^{odd}(\bq,E)=
&\frac{1}{2\pi^2}
\int_{\bk} {\rm Tr}\Bigl[\tau_3
\int dE'~\bigl[\frac{1}{E-E'}\sum_\bk  G''_{\bk-}(E){\hat t}(\bq,\bk)
G''_{\bk+}(E')-[E\leftrightarrow E']\bigr]
\Bigr].
\end{eqnarray}
As we introduce  the joint density of states, 
\begin{eqnarray}\label{JDOSgen}
&J(\bq,E,E')
=\frac{1}{\pi^2}\sum_\bk Tr[\tau_3 G''_{\bk-}(E){\hat t}(\bq,\bk)
G''_{\bk+}(E')],
\end{eqnarray}
(\ref{deltarhooddTJ}) becomes
\begin{eqnarray}\label{deltarhooddTJgen}
&\delta \rho^{odd}(\bq,\omega)=
&\frac{1}{2}
\int dE'~\frac{1}{E-E'}[J(\bq,E,E')+J(\bq,E',E)].
\end{eqnarray}
The Fourier transformed conductance ratio $Z(\bq,E)$ given by (\ref{Zphsym})
now becomes (for $\bq\neq 0$)
\begin{eqnarray}\label{Z2}
&Z(\bq,E) 
= \frac{1}{\rho_0(E)}\int dE'~\frac{1}{E-E'}[J(\bq,E,E')+J(\bq,E',E)].
\end{eqnarray}
Substituting the expression for the BCS Green's function (\ref{GBCS}) in
(\ref{JDOSgen}),
we obtain  \begin{eqnarray}\label{JDOSBCS}
J(\bq,E,E')
=&\frac{1}{4}\sum_\bk \frac{1}{E_{\bk+}E_{\bk-}}
Tr[\tau_3 (E+\epsilon_{\bk-}\tau_3+\Delta_{\bk-}\tau_1){\hat t}(\bq,\bk)
(E'+\epsilon_{\bk+}\tau_3+\Delta_{\bk+}\tau_1)]\cdot\nonumber\\
\cdot& [\delta(E-E_{\bk-})-
\delta(E+E_{\bk-})]
[\delta(E'-E_{\bk+})-
\delta(E'+E_{\bk+})]
\cdot sgn E\cdot sgn E',
\end{eqnarray}
where 
$E_{\bk_\pm}\equiv\sqrt{\epsilon^2_{\bk_\pm}+\Delta^2_{\bk_\pm}}$. 
Provided both the energies are positive, $E,E'>0$,
we obtain \begin{eqnarray}\label{JDOS}
&J(\bq,E,E')
=\sum_{\bk_1,\bk_2} C(\bk_1,\bk_2)\delta(E- E_{\bk_1})\delta(E'-E_{\bk_2})\delta^{(2)}
(\bk_1-\bk_2-\bq),
\end{eqnarray}
where the coherence factor is
\begin{eqnarray}\label{C}
C(\bk_1,\bk_2)\equiv \frac{1}{4}
\frac{1}{E_{\bk_1}E_{\bk_2}}
Tr[\tau_3 (E+\epsilon_{\bk_1}\tau_3+\Delta_{\bk_1}\tau_1){\hat t}(\bk_1,\bk_2)
(E'+\epsilon_{\bk_2}\tau_3+\Delta_{\bk_2}\tau_1)].
\end{eqnarray}
Now the fluctuations in the conductance ratio at wavevector $\bq$
are given by:
\begin{eqnarray}\label{Z-JDOS}
Z(\bq,E)|_{{\bq\neq 0}}\propto 
\int \frac{dE'}{E-E'}\int d\bk_1d\bk_2C(\bk_1,\bk_2)
\delta(E- E_{\bk_1})\delta(E'-\beta E_{\bk_2})\delta^{(2)}
(\bk_1-\bk_2-\bq).
\end{eqnarray}
Thus, the fluctuations in the conductance ratio $Z(\bq,E)$  are
determined by
a Kramers-Kronig transform of the joint density of states with a well-defined
coherence factor.


Conventionally, coherence factors
appear in dissipative responses, such as (\ref{JDOS}). The
appearance of a Kramers-Kronig transform reflects the fact that tunneling
conductance is determined by the non-dissipative component of the
scattering. 
The validity of the octet model depends on the presence of sharp peaks
in the joint density of states. We now argue that 
if the joint density of states contains
sharp peaks at well-defined points in momentum space, then these peaks survive through
the Kramers-Kronig procedure, so that they still 
appear in the conductance ratio $Z(\bq,E)$
with a non-Lorentzian profile, but precisely the
 same coherence factors. We can illustrate this point both
numerically and  analytically. Fig. 1 contrasts joint density of
states with the Fourier transformed conductance ratio $Z(\bq,E)$ for scattering
off a weak scalar impurity, showing the appearance of the ``octet''
scattering wavevectors in both plots. Similar comparisons have been
made by earlier authors\cite{Pereg-Barnea-Franz,Scalapino}. 
  
Let us now repeat this analysis analytically.   
Suppose $J(\bq,E_1,E_2)$ (\ref{JDOS})
has a sharp peak at an octet $\bf{q}$ vector, $\bq=\bq_i$
($i=1-7$), defined by the delta function
$J(\bq,E_1=E,E_2=E)=C_i\delta^{(2)}(\bq-\bq_i)$, where  $C_i$
is the energy-dependent coherence factor for the $i$th octet
scattering process. When we vary the 
energy $E_2$ away from $E$,
the position of the 
characteristic octet vector will drift according to 
\begin{eqnarray}
&\bq_i(E_1,E_2)=\bq_i(E)-\nabla_{E_1}\bq_i(E_1-E)+\nabla_{E_2}\bq_i(E_2-E),
\end{eqnarray}
where $\nabla_{E_1}\bq_i=\frac{
1}{v_\Delta}\hat{{\bf n}}_1 (i)
$ and  $\nabla_{E_2}\bq_i=\frac{1}{v_\Delta}\hat{{\bf n}}_2 (i)$ are 
directed along the initial and final quasiparticle velocities, and $v_\Delta$ is the 
 quasiparticle group velocity. 
Carrying out the integral over $E'$ in (\ref{Z-JDOS}) we now obtain 
\begin{eqnarray}\label{Z-exp}
Z(\bq,E)&\propto& 
\int dE'~\frac{C_i}{E-E'}\Bigl[
\delta(\bq-\bq_i(E)-\frac{\hat{n_2}}{v_\Delta}(E'-E))+
\delta
(\bq-\bq_i(E)+\frac{\hat{n_1}}{v_\Delta}(E'-E))\Bigr]\nonumber
\\
&=&C_i\Bigl[ \frac{1}{(\bq-\bq_i)_{\| 1}}\delta\bigl((\bq-\bq_i)_{\perp 1}
\bigr )-
\frac{1}{(\bq-\bq_i)_{\| 2}}\delta\bigl ((\bq-\bq_i)_{\perp 2}
\bigr )\Bigr],
\end{eqnarray}
where
\[
(\bq -\bq_{i})_{\| 1,2}
= (\bq -\bq_{i})\cdot \hat {\bf n}_{1,2} (i)
\]
denotes the component of $(\bq -\bq_{i})$ parallel to the
initial/final quasiparticle velocity and 
\[
(\bq -\bq_{i})_{\perp 1,2}
= (\bq -\bq_{i})\cdot[{\hat {\bf z}}
\times \hat {\bf n}_{1,2} (i)]\cdot 
\]
denotes the component of $(\bq -\bq_{i})$ {\sl perpendicular} to the
initial/final quasiparticle velocity, where $\hat {\bf{z}}$ is the
normal to the plane.
Thus, a single sharp peak in the joint density of states produces an
enhanced dipolar distribution in the conductance ratio $Z(\bq,E)$,
with the axes of the dipoles aligned along the directions of the
initial and final quasiparticle velocities. The above analysis can be
further refined by considering the Lorentzian distribution of the
quasiparticle interference peaks, with the same qualitative
conclusions.

  To summarize, the conductance ratio $Z(\bq,E)$ is a spectral probe for
  fluctuations in the quasiparticle charge density in response to disorder.
  $Z(\bq,E)$ is characterized by the joint coherence factors of charge 
  ($\tau_3$) and the scattering potential. Provided the original  joint density
  of states is sharply peaked at the octet vectors $\bq_i,~i=1-7$, the conductance ratio $Z(\bq,E)$ 
  is also peaked   at the octet vectors $\bq_i,~i=1-7$.
  \begin{figure}
     \centering
     \subfigure[]
     {
          \label{jdos}
\includegraphics[width=0.45\linewidth]{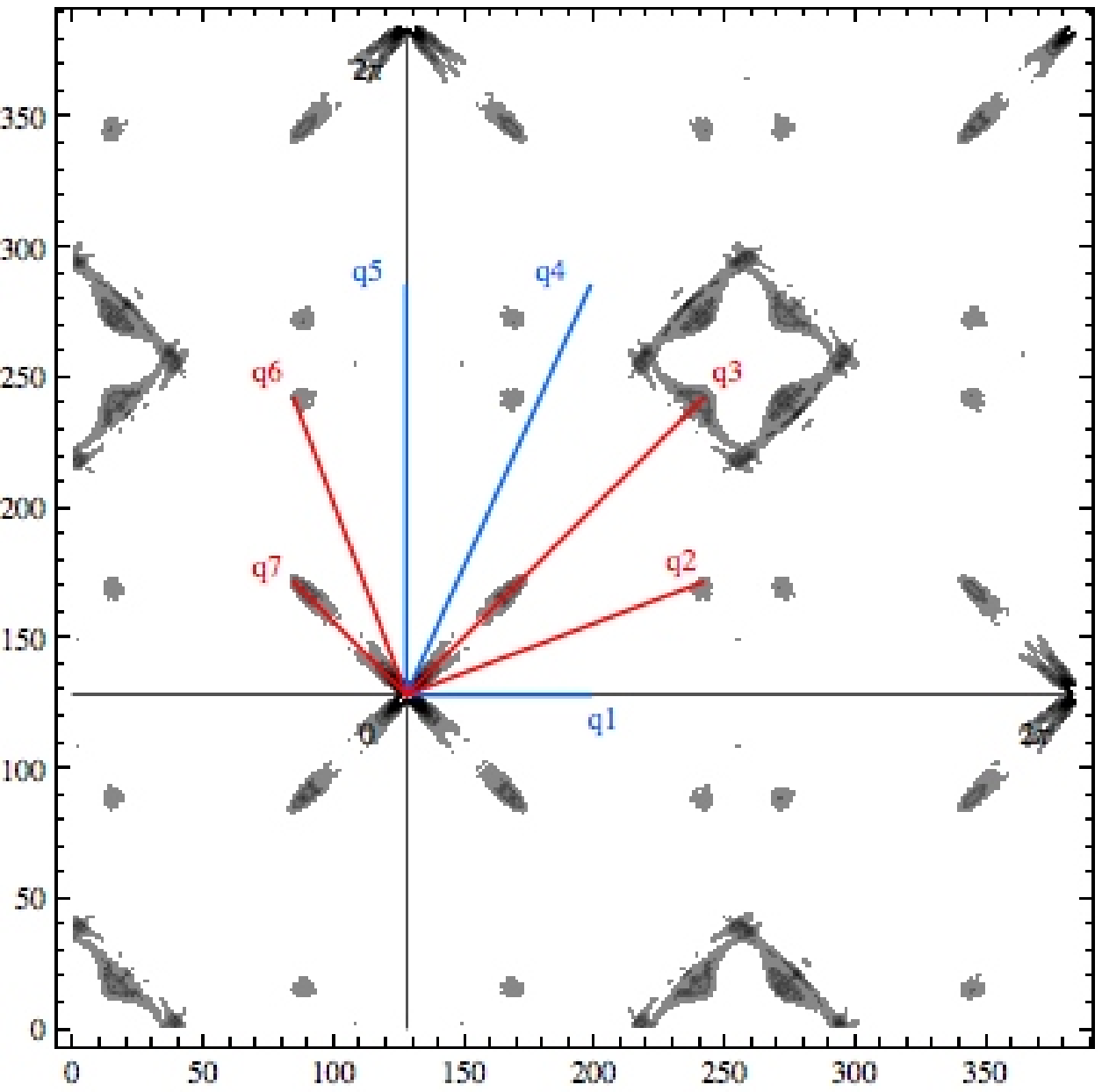}
     }
     \hspace{.3in}
    \subfigure[]
     {
       \label{sca}
                
          \includegraphics[width=0.45\linewidth]{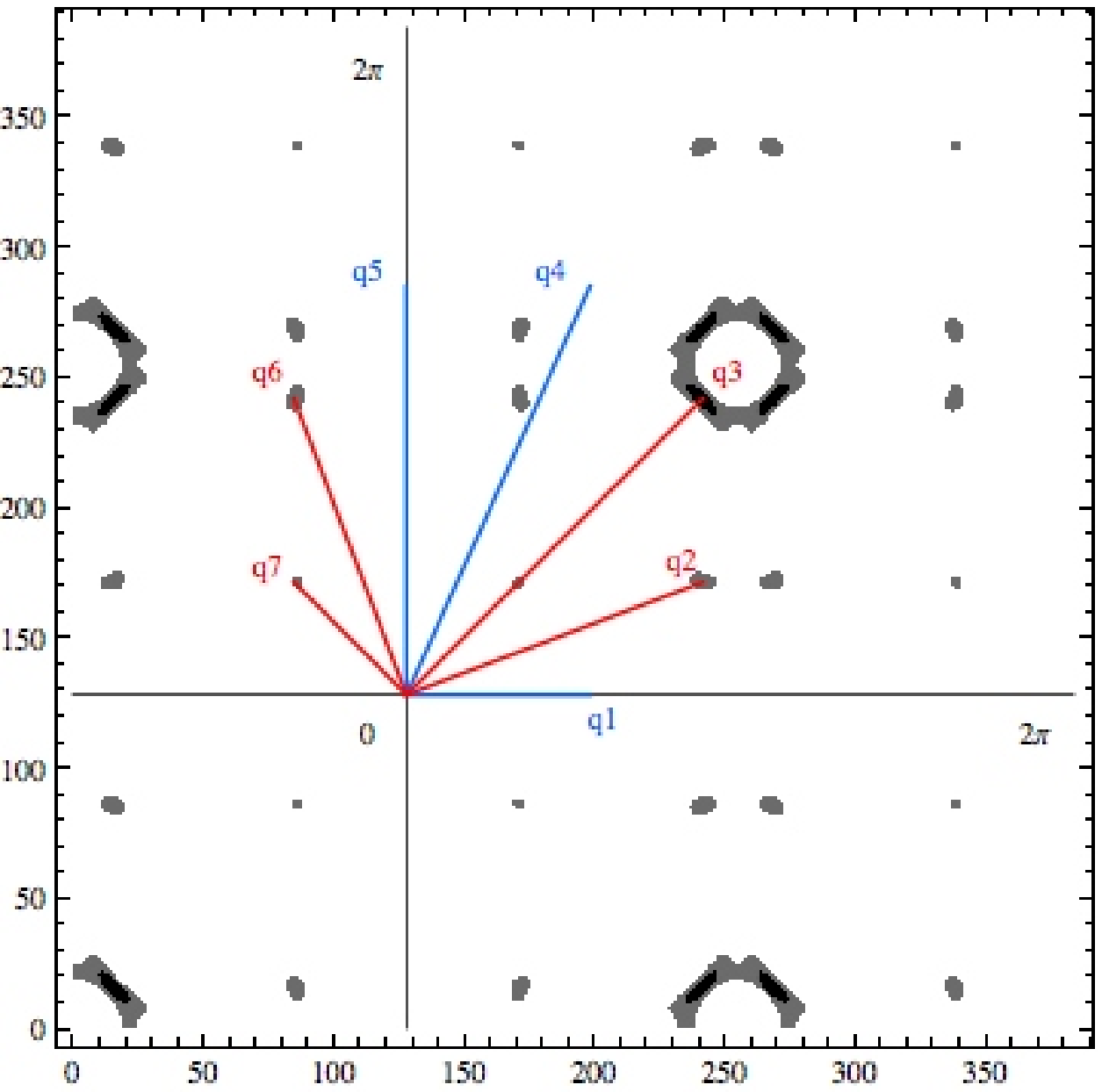}
	  }
	  
	 \caption{(Color online) Observation of coherence factor effects in the squared 
	 joint density of states  $|J(\bq,V,V)|^2$ and in the squared Fourier transformed
	 conductance ratio $|Z(\bq,V)|^2$.
	 Fig. (a) shows the squared joint density of states $|J(\bq,V,V)|^2$
	 at the bias voltage $V=\Delta_0/2$,
	 Fig. (b) shows the squared Fourier transformed conductance ratio $|Z(\bq,V)|^2$
	 produced by
a weak scalar  scattering potential ${\hat t}(\bq)={\hat\tau_3} $.
Red lines label the positions of
the sign-reversing q-vectors $q= q_{2,3,6,7}$, where weak scalar 
 scattering  is peaked.
Blue lines label the positions of the
sign-preserving q-vectors $q= q_{1,4,5}$, where  weak scalar 
 scattering  is minimal.}
\label{J-scalar}
\end{figure}

\section{Model for quasiparticle interference in vortex lattice}

Next, we discuss the recent experiments by Hanaguri et al.\cite{Hanaguri}
on the underdoped cuprate superconductor calcium oxychloride,
$\ca$ (Na-CCOC), which have successfully observed the coherence factor effects with
 Fourier Transform Scanning Tunneling Spectroscopy
(FT-STS)  in a magnetic field.  The main observations are:

\begin{itemize}
\item  {\bf{ A selective enhancement of sign-preserving; depression of 
sign-reversing scattering events}}. 
In a field, Hanaguri et al.\cite{Hanaguri} observe a selective
enhancement of the scattering events between parts of the Brillouin zone with
the same gap sign, and  a selective depression of the scattering events
between parts of the Brillouin zone with opposite gap signs,
 so that the sign-preserving q-vectors $\bq_{1,4,5}$
are enhanced, and the  sign-reversing q-vectors $\bq_{2,3,6,7}$ are depressed.

\item {\bf{Large  vortex cores}} with a core size $\xi\sim 10a$ of
order ten lattice constants. Experimentally, vortex cores are imaged
as regions of shallow gap \cite{Hanaguri}.  The figure $\xi\sim 10a $ is consistent with
magnetization and angular resolved photoemission (ARPES) measurements \cite{largecores}.

\item {\bf High momentum transfer scattering} involving momentum transfer 
over  a large fraction of the Brillouin zone size at $q_{4,5}\sim k_F$.
A paradoxical feature
of the observations is the enhancement of high momentum 
transfer $q\sim  \pi/a $ scattering by objects  
that are of order ten lattice spacings in diameter. 
The enhanced high momentum scattering 
clearly reflects sub-structure on length scales much smaller than the
vortex cores. 

\item {\bf Core-sensitivity}. 
Fourier mask analysis reveals that the 
scattering outside the vortex core regions differs qualitatively from
scattering inside the vortex core regions.  In particular, the enhancement
of the sign-preserving scattering events is associated with the signal
inside the ``vortex cores'', whereas the depression of the
sign-reversing scattering events is mainly located outside the vortex
regions.

\end{itemize}

Recently, T. Pereg-Barnea and M. Franz \cite{Pereg-Barnea-Franz2}
have proposed an initial interpretation of these observations in terms of quasiparticle
scattering off vortex cores. Their model explains the enhancement of the sign
preserving scattering in the magnetic field in terms of scattering off vortex
cores, provided vortex cores are small 
with $\xi\sim a$, as in high temperature superconductor
$Bi_2Sr_2CaCu_2O_{8+\delta}$ (Bi2212). However, the large vortex core size of $\ca$ is unable to account
for the field-driven enhancement in the high momentum scattering.

Motivated by this observation, we have developed an alternative
phenomenological model to interprete the high-momentum scattering. 
In our model, vortices bind to individual impurities,
incorporating them into their cores and modifying their scattering
potentials. 
This process replaces random potential scattering off the original impurities
with gap-sign-preserving Andreev reflections off order parameter modulations in
the vicinity of the pinned vortices.  The high-momentum transfer scattering, involved in the selective enhancement and
suppression, originates from the impurities whose
scattering potentials are modified by the presence of the vortex lattice.
Rather than attempt a detailed microscopic model for the pseudo-gap state
inside the vortex cores and impurities bound therein, our approach attempts to
characterize the scattering in terms of phenomenological form factors
that can be measured and extracted from the data.

\subsection{Construction of the model}

In the absence of a field, random fluctuations in the tunneling
density of states are produced by the original impurities. 
We assume that scattering off the impurities is mutually independent
permitting us to write the change in density of states as a sum of
contributions from each impurity
\begin{equation}\label{B0}
\delta \rho ({\br},V,B=0) =
 \sum_j \delta \rho_{i} ({\br-\br_j},V),
\end{equation}    
where $\br_{j}$ denote the positions of the impurities. If 
 \[
n_{i}= \hbox{original concentration of impurities in the absence of magnetic field},
\]
then we obtain
\begin{eqnarray}\label{corrB0}
R(\bq,V,B=0)= &n_{i}~\overline{ {\delta \rho_{i} ({\bq},V)
\delta
\rho^*_{i}({-\bq},V)}}.
\end{eqnarray}

Next we consider how  the quasiparticle
scattering changes in the presence  of a magnetic field.
Pinned
vortices arising in the magnetic field act as new scatterers. In the experiment
\cite{Hanaguri}, vortices are pinned to the preexisting disorder, so that
in the presence of a magnetic field, there are essentially three types
of scatterers:

\begin{itemize}

\item bare impurities,
\item vortices,  
\item vortex-decorated impurities.

\end{itemize}
Vortex-decorated impurities are impurities lying 
within a coherence length of the center of a vortex core. 
We assume that these three types of scattering centers act as independent
scatterers, so
that the random variations in the tunneling density of states are given by the sum of the independent contributions,
from each type of scattering center:
\begin{equation}
\delta \rho ({\br},V,B) =
\sum_j \delta \rho_{V} ({\br-\br_j},V)+
 \sum_l \delta \rho_{DI} ({\br-\br'_l},V)+
 \sum_m \delta \rho_{I} ({\br-\br''_m},V),
\end{equation}    
where $\br_{j},\br'_{l},\br''_{m}$ denote the positions of vortices, decorated
impurities and bare impurities, respectively. 
In a magnetic field, the concentration of vortices  is given by
\[
n_{V}= \hbox{concentration of vortices}=\frac{2eB}{h},
\]
In each vortex core, there will be $n_{core}= n_{i}\pi (\xi^{2}/4)$ 
impurities, where $\pi (\xi^2)/4$ is the  area of a vortex
and $n_{i}$ is the original concentration of bare scattering centers
in the absence of a field. 
The concentration of vortex-decorated impurities is then given by
\[
n_{DI}= \hbox{concentration of vortex-decorated
impurities}=n_{core}n_{V}
=\frac{2eB}{h}n_i\pi (\xi/2)^2.
\]
Finally, the residual concentration of ``bare'' scattering
centers is given by
\begin{equation}\label{n_BI}
n_{I}= n_{i} - n_{DI} = \hbox{concentration of residual ``bare'' impurities}.
\end{equation}
Treating the three types of scatterers as independent, we  write
\begin{eqnarray}\label{corrB}
R(\bq,V)= &
n_{V}~ \overline{{\delta \rho_{V} ({\bq},V)\delta \rho_{V}^{*}
({-\bq},V)}}
+n_{DI} ~\overline{ {\delta \rho_{DI}({\bq},V)\delta
\rho^{*}_{DI}({-\bq},V)}}\nonumber\\
+&\left(n_{i}-n_{DI} \right)~\overline{ {\delta \rho_{I}({\bq},V)\delta
\rho^{*}_{I}({-\bq},V)}}.
\end{eqnarray}
The first term in (\ref{corrB}) accounts for the  quasiparticle scattering off
the vortices, the second term
accounts for the quasiparticle scattering off the vortex-decorated
impurities and
the third term
accounts for the quasiparticle scattering off the residual bare impurities in the presence of the superflow.
It follows that 
\begin{equation}\label{Zfull}
|Z(q,V,B)|^{2} = \frac{2eB}{h}~|Z_{V}(q,V,B)|^{2}+
\frac{2eB}{h}n_{core}|Z_{DI}(q,V,B)|^{2}+
(n_i - \frac{2eB}{h}n_{core})~
|Z_{I}(q,V,B)|^{2},
\end{equation}
where   $Z(q,V,B)$ is given by (\ref{Zphsym}), averaged over the vortex
configurations,  $Z_{V} (q,V,B)$, $Z_{DI}(q,V)$ and $Z_{I}(q,V)$ are Fourier images of the Friedel oscillations in the tunneling density
of states induced by the vortices, vortex-decorated impurities
and the bare impurities in the presence of the superflow.
Our goal here is to model the quasiparticle
scattering phenomenologically, without a recourse
to a specific microscopic model of the scattering in the vortex interior.
To achieve this goal, we introduce 
$Z_{VI}(q,V,B)$, a joint conductance ratio of the vortex-impurity composite,
which encompasses the scattering off a vortex core and the impurities
decorated by the vortex core,
\begin{eqnarray}\label{ZVI}
|Z_{VI}|^2=|Z_{V}|^2+n_{core}|Z_{DI}|^2,
\end{eqnarray}
so that we obtain
\begin{equation}\label{Z(q,V,B)}
|Z(q,V,B)|^{2} = \frac{2eB}{h}~|Z_{VI}(q,V,B)|^{2}+
 (n_i - \frac{2eB}{h}n_{core})
|Z_{I}(q,V)|^{2}.
\end{equation}
This expression describes quasiparticle scattering in a clean superconductor
in low magnetic fields in a model-agnostic way, namely, it is valid regardless of
the choice of the detailed model
of quasiparticle scattering in the vortex region. $Z_{VI}(q,V,B)$
here describes the scattering  off the vortex-impurity composites, which we now
proceed to discuss. 

\subsection{Impurities inside the vortex core: calculating  $Z_{VI}$}
As observed in the  conductance
ratio $Z(\bq,V,B)$, the intensity of scattering  between parts of
the Brillouin zone with the same sign of the gap 
grows in the magnetic field, which implies that 
the scattering potential of a vortex-impurity composite has a predominantly
sign-preserving coherence factor.


We now turn to a discussion of the scattering mechanisms that can
enhance sign-preserving scattering inside the vortex cores.
Table 1 shows a list of
scattering potentials and their corresponding coherence factor effects.
Weak potential scattering is immediately excluded.  Weak scattering off
magnetic impurities can also be excluded, since the change in the
density of states of the up and down electrons cancels. This leaves
two remaining contenders: Andreev scattering off a
fluctuation in the gap function, and multiple scattering, which
generates a t-matrix proportional to the unit matrix.

We can, in fact, envisage both scattering mechanisms being active in
the vortex core. Take first the case of a resonant scattering center. 
In the bulk superconductor, the effects of a
resonant scatterer are severely modified by the presence of the
superconducting gap \cite{Balatsky}. When the same scattering
center is located inside the vortex core where the superconducting
order parameter is depressed, we envisage that the resonant scattering
will now be enhanced.

On the other hand, we can not rule out Andreev scattering.  A scalar
impurity in a d-wave superconductor scatters the gapless
quasiparticles, giving rise to Friedel oscillations in the order
parameter that act as Andreev scattering centers \cite{Nunner,Pereg-Barnea-Franz,Pereg-Barnea-Franz2}.  Without a detailed
model for the nature of the vortex scattering region, we can not say
whether this type of scattering is enhanced by embedding the impurity
inside the vortex. For example, if, as some authors have suggested
\cite{WignerSuperSolid}, the competing pseudo-gap
phase is a Wigner supersolid, then the presence of an impurity may
lead to enhanced oscillations in the superconducting order parameter
inside the vortex core.

With these considerations in mind, we consider both sources of
scattering as follows
\begin{equation}\label{}
\hat t (\bq ,\bk ,i\omega_n) 
= 
t_{A}(\bq ,\bk ,i\omega_n) +
t_{R}(\bq ,\bk ,i\omega_n)
\end{equation}
where 
\begin{align*}
\hat{t}_A(\bq,\bk,i\omega_n)=
 \frac{1}{2}\Delta_0
f_A(\bq)(\chi_{\bk_+}+\chi_{\bk_-})\hat{\mbox{\boldmath{$\tau$}}}_1,\qquad \qquad (\hbox{Andreev scattering})
\end{align*}
describes the Andreev scattering. Here
 $\chi_\bk=c_x-c_y$ is the d-wave function with $c_{x,y}\equiv\cos k_{x,y}$.
The resonant scattering is described by
\[
{\hat t}_R(\bq,\bk ,i\omega_n)=i\Delta_{0}  \hbox{sgn}
(\omega_n)~f_R(\bq){\bf 1}. \qquad \qquad (\hbox{Resonant scattering})
\]

Using the T-matrix approximation, we obtain for the even and odd components of
Fourier transformed  fluctuations in the local density of states due to
the scattering off the superconducting order parameter amplitude modulation,
\begin{eqnarray}\label{deltarhoeven-oddV}
&\delta \rho^{even}_{VI}(\bq,\omega)=\frac{1}{2\pi }
&{\rm Im} \int_{\bk} {\rm Tr}\Bigl[
{G}_{\bk_{-}} (\omega-i\delta )\ \hat{t}(\bq,\bk,\omega-i\delta )
\ { G}_{\bk_{+}} (\omega-i\delta )
\Bigr],\\
&\delta \rho^{odd}_{VI}(\bq,\omega)=\frac{1}{2\pi }
&{\rm Im} \int_{\bk} {\rm Tr}\Bigl[\tau_3
{G}_{\bk_{-}} (\omega-i\delta )\ \hat{t}(\bq,\bk,\omega-i\delta )
\ { G}_{\bk_{+}} (\omega-i\delta )
\Bigr],
\end{eqnarray}
where $\bk_{\pm }= \bk  \pm \bq /2$, $G_{\bk } (\omega)= [\omega
-\epsilon_{\bk }\tau_{3}-\Delta_{\bk }\tau_{1}]^{-1}$ is the Nambu
Green's function for an electron with normal state dispersion
$\epsilon_{\bk }$ and gap function $\Delta_{\bk }$. 
We now obtain 
\[
\delta \rho^{even(odd)}_{V}(\bq ,\omega)=
f_A(\bq)\Lambda_A^{even(odd)}(\bq ,\omega)+
f_R(\bq)\Lambda_R^{even(odd)}(\bq ,\omega)
\]
with
\begin{eqnarray}\label{Lambda-even}
\Lambda^{even}_{A} (\bq ,\omega)&=& \frac{\Delta_{0}
}{4\pi }{\rm Im}\int_k~
 (\chi_{\bk_+}+\chi_{\bk_-})~\Bigl[
\frac{z(\Delta_{\bk_+}+\Delta_{\bk_-})}
{(z^2-E_{\bk_+}^2)
(z^2-E_{\bk_-}^2)}\Bigr]_{z=\omega-i\delta },\\
\Lambda^{even}_{R} (\bq ,\omega)&=& \frac{\Delta_{0}
}{2\pi }{\rm Im}\int_k~
~\Bigl[
\frac{-i(z^2+\epsilon_{\bk_+}\epsilon_{\bk_-}+\Delta_{\bk_+}
\Delta_{\bk_-})}
{(z^2-E_{\bk_+}^2)
(z^2-E_{\bk_-}^2)}\Bigr]_{z=\omega-i\delta }.
\end{eqnarray}
The substantially smaller odd components are:
\begin{eqnarray}\label{Lambda-odd}
\Lambda^{odd}_{A} (\bq ,\omega)&=& \frac{\Delta_{0}
}{4\pi }{\rm Im}\int_k~
 (\chi_{\bk_+}+\chi_{\bk_-})~\Bigl[
\frac{\epsilon_{\bk_+}\Delta_{\bk_-}+\epsilon_{\bk_-}\Delta_{\bk_+}}
{(z^2-E_{\bk_+}^2)
(z^2-E_{\bk_-}^2)}\Bigr]_{z=\omega-i\delta },\\
\Lambda^{odd}_{R} (\bq ,\omega)&=& \frac{\Delta_{0}
}{2\pi }{\rm Im}\int_k~
~\Bigl[
\frac{-i~z(\epsilon_{\bk_+}+\epsilon_{\bk_-})}
{(z^2-E_{\bk_+}^2)
(z^2-E_{\bk_-}^2)}\Bigr]_{z=\omega-i\delta }.
\end{eqnarray} 
where $E_{\bk}=[\epsilon_{\bk}^2+\Delta_{\bk}^2]^{\frac{1}{2}}$ is the
quasiparticle energy. 
The
vortex contribution to
the Fourier transformed conductance ratio (\ref{Z(q,V,B)}) is then
\begin{eqnarray}\label{Z_VI}
Z_{VI} (\bq ,V,B) = n_{V}( Z_{A} (\bq ,V,B)+ Z_{R} (\bq ,V,B)),
\end{eqnarray}
where
\begin{eqnarray}\label{Z_A}
Z_{A} (\bq ,V,B)=f_A(\bq)~\biggl[
(\frac{1}{\rho_0(V)}-\frac{1}{\rho_0(-V)})\Lambda_A
^{even}(\bq ,V)+(\frac{1}{\rho_0(V)}+\frac{1}{\rho_0(-V)})\Lambda_A
^{odd}(\bq ,V)\biggr]
\end{eqnarray}
and
\begin{eqnarray}\label{Z_R}
Z_{R} (\bq ,V,B)=f_R(\bq)~\biggl[
(\frac{1}{\rho_0(V)}-\frac{1}{\rho_0(-V)})\Lambda_R
^{even}(\bq ,V)+(\frac{1}{\rho_0(V)}+\frac{1}{\rho_0(-V)})\Lambda_R
^{odd}(\bq ,V)\biggr].
\end{eqnarray}
\section{Numerical simulation}

In this section we
compare the results of our phenomenological model with the
experimental data by numerically computing 
$Z_{VI}(\bq,V,B)$ (\ref{Z_VI})
for Andreev  (\ref{Z_A}) and resonant (\ref{Z_R}) scattering.

In these calculations we took  a BCS superconductor with a d-wave gap $\Delta_{\bk}=\Delta_0/2(\cos k_x-\cos k_y)$ with
$\Delta_0=0.2 t$
and  a dispersion which has been introduced to fit the Fermi surface
of an underdoped $\ca$ sample with $x=0.12$ \cite{ShenThesis}:
\[
\epsilon_\bk=-2t(\cos k_x+\cos k_y)-4t'\cos k_x\cos k_y-
2t''(\cos 2k_x+\cos 2k_y)+\mu,
\]
where  $t=1$, $t'=-0.227$, $t''=0.168$,  $\mu=0.486$.

\subsection{Evaluation of $Z_{VI}$}

In the absence of a microscopic model for the  interior of the
vortex core, we model the Andreev  and the resonant  scattering 
 in the vortex region by constants 
$f_A (\bq,i\omega_n )=f_A$   and $f_R (\bq,i\omega_n )=f_R$. 
Fig. 2 shows the results of calculations using these assumptions.

Our simple model reproduces the enhancement of sign-preserving q-vectors
$q_{1,4,5}$ as a result of Andreev  and  resonant scattering off
vortex-impurity composites. 
Some care is required in interpreting Fig. 2, because the squared conductance
ratio $|Z(\bq ,V)|^{2}$
contains weighted contributions from both even and odd fluctuations in
the density of states, with the weighting factor favoring {\sl odd}
fluctuations, especially near $V=0$. 
Both Andreev and resonant
scattering contribute predominantly to the
even fluctuations of the density of states (see Table 1), and give rise to the
signals at $q_{1,4,5}$. In the case of resonant scattering, we observe
an additional 
peak at $q_{3}$. 
From Table 1, we see that the Andreev and the resonant 
scattering potentials also produce a signal in the odd channel which
experiences no coherence factor effect, contributing to all the octet
q-vectors, which, however, enters the conductance ratio $Z(\bq ,V)$ given by
(\ref{Z-even-odd})
with a substantial weighting factor.
This is the origin of the peak at $q_ {3}$ in Fig. 2(b).

\begin{figure}
     \centering
     \subfigure[]
     {
          \label{Andr}
               
\includegraphics[width=0.45\linewidth]{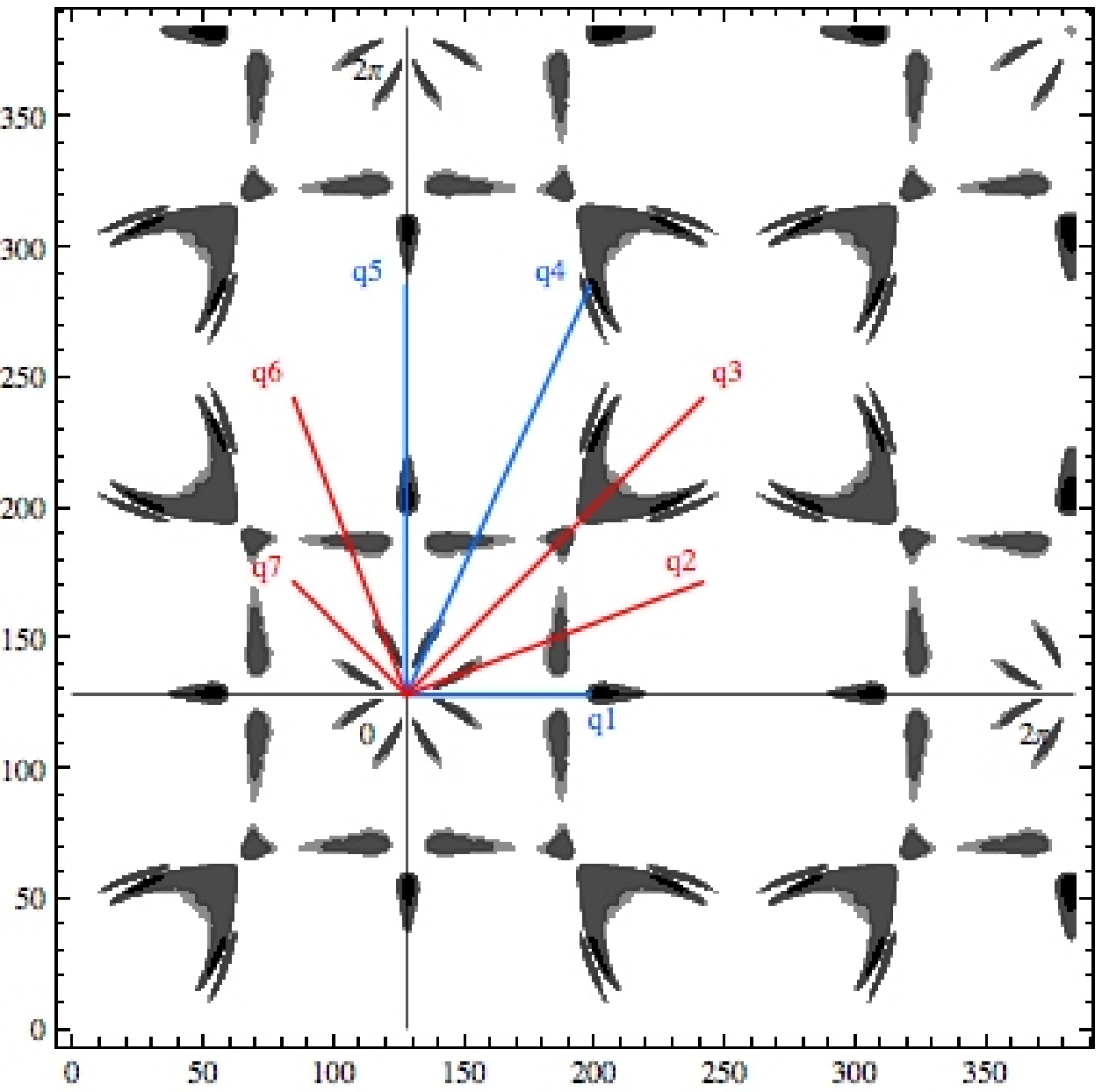}
     }
     \hspace{.3in}
    \subfigure[]
     {
       \label{Res}
          \includegraphics[width=0.45\linewidth]{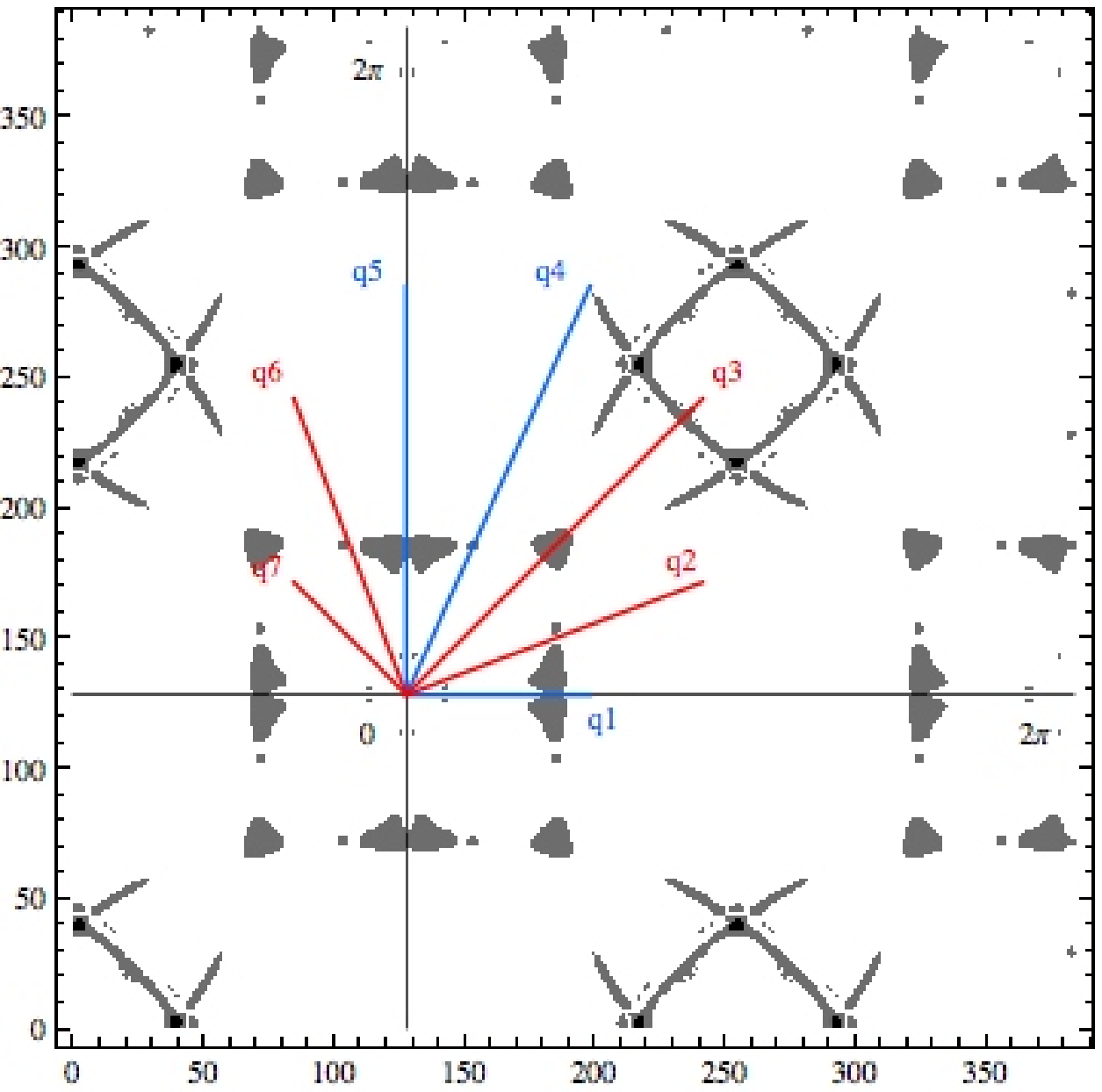}
	  }
	  
	  \caption{(Color online) Quasiparticle interference produced by
the Andreev and the resonant scattering potentials, the primary candidates for
producing the experimentally observed enhancement of
sign-preserving scattering. Fig. (a) displays a density plot of the
squared Fourier transformed conductance ratio $|Z_{A}(\bq,V)|^2$ 
predicted by (\ref{Z_A}) at a bias voltage $V= \Delta_{0}/2$ produced
by pure Andreev scattering ($f_{A}\neq 0$, $f_{R}=0$).
Fig. (b) displays a density plot of the
squared Fourier transformed conductance ratio $|Z_{R}(\bq,V)|^2$ 
predicted by (\ref{Z_R}) at a bias voltage $V= \Delta_{0}/2$ produced by
resonant scattering ($f_{R}\neq 0$, $f_{A}=0$).
Blue lines label the positions of the
sign-preserving q-vectors $q= q_{1,4,5}$, where both Andreev and resonant
scattering is peaked.
Red lines label the positions of
the sign-reversing q-vectors $q= q_{2,3,6,7}$, where  both Andreev and resonant
scattering  is minimal.}
\end{figure}

\subsection{Comparison with experimental data}

The results of the calculation of the full squared conductance ratio
$|Z(q,V,B)|^2$ are obtained by combining the scattering off the
impurities inside the vortex core $Z_{VI}$ with the contribution from
scattering off impurities outside the vortex core  $Z_{I}$, according
to equation (\ref{Z(q,V,B)}), reproduced here:
\begin{equation}\label{Z(q,V,B)again}
|Z(q,V,B)|^{2} = \frac{2eB}{h}~|Z_{VI}(q,V,B)|^{2}+
 (n_i - \frac{2eB}{h}n_{core})~
|Z_{I}(q,V,B)|^{2}.
\end{equation}
where $n_{core}= n_{i}\pi (\xi/2)^{2}$ is the number of impurities per
vortex core.
Fig. 3 displays a histogram of the 
computed field-induced change in the conductance
ratio  $|Z(\bq_{i},V,B)|^2-|Z(\bq_{i},V)|^2$ at the octet
q-vectors.
In these calculations, we took an equal strength of Andreev
and resonant scattering $f_{R}=f_{A}$, with a weak scalar scattering
outside the vortex core of strength $f_{I}= f_{R}=f_{A}$.
In all
our calculations, we find that Andreev and resonant  scattering are
equally effective in qualitatively modelling the observations.
The main effect governing the depression of 
sign-preserving wavevectors $q_{1,4,5}$ derives from the change in the
impurity scattering potential that results from embedding the impurity
inside the vortex core.  

We estimated the percentage of the impurities decorated by the vortices
from the fraction of sample area covered by the vortices. 
The concentration of vortices is $n_V(B)=2eB/h=B/\Phi_0$, where 
$\Phi_0=h/ (2e)=2.07\times 10^{-15} $ weber is the superconducting
magnetic flux quantum.
The area of a vortex region is estimated as 
$A_V=\pi(\xi_0/2)^2$ with the superconducting coherence length
$\xi_0=44$ \AA \cite{Kim},
so that the percentage of the original impurities that are decorated by vortices in the presence of the magnetic field is 
 $\alpha(B)=n_V(B)~A_V$. Using these values, 
 we obtain for the magnetic field of $B=5$ T $\alpha(B=5~T)\approx 3.7\%$, and for
 $B=11$ T  $\alpha(B=11~T)\approx 8.1\%$.
For simplicity, we assume that a vortex core is pinned to a single impurity,
$n_{core}= n_{i}\pi (\xi/2)^{2}=1$, so that the  
ratio of the concentrations of the impurities and vortices is
$n_i/n_V(B)~=n_{core}/A_V/(2eB/h)$, which becomes for $B=5$ T 
$n_i/~n_V(B=5T)\approx 27$, and for $B=11$ T 
$n_i/n_V(B=11T)\approx 12$. 

\begin{figure}
     \centering
     \subfigure[]
     {
       \label{histT-Andr}
                
          \includegraphics[width=0.45\linewidth]{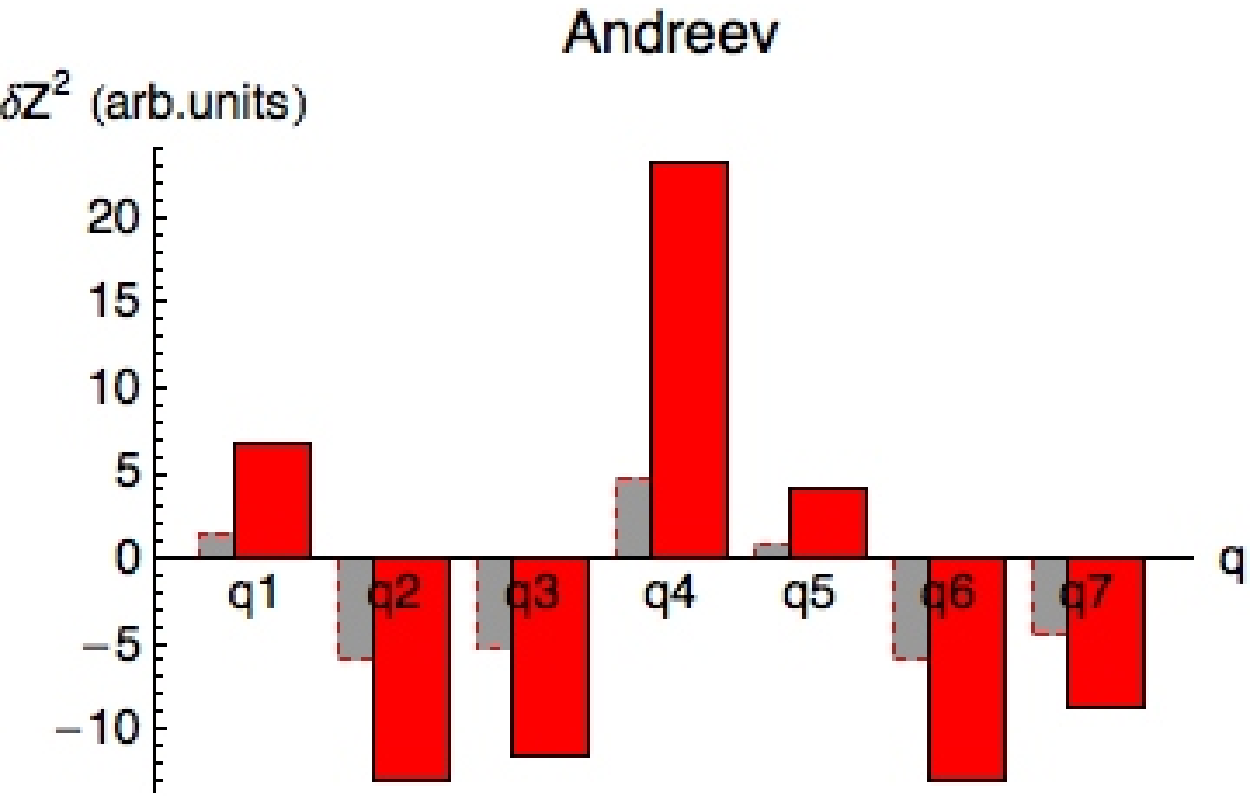}
	  }
	  \hspace{.3in}
    \subfigure[]
     {
       \label{histT-Res}
                
          \includegraphics[width=0.45\linewidth]{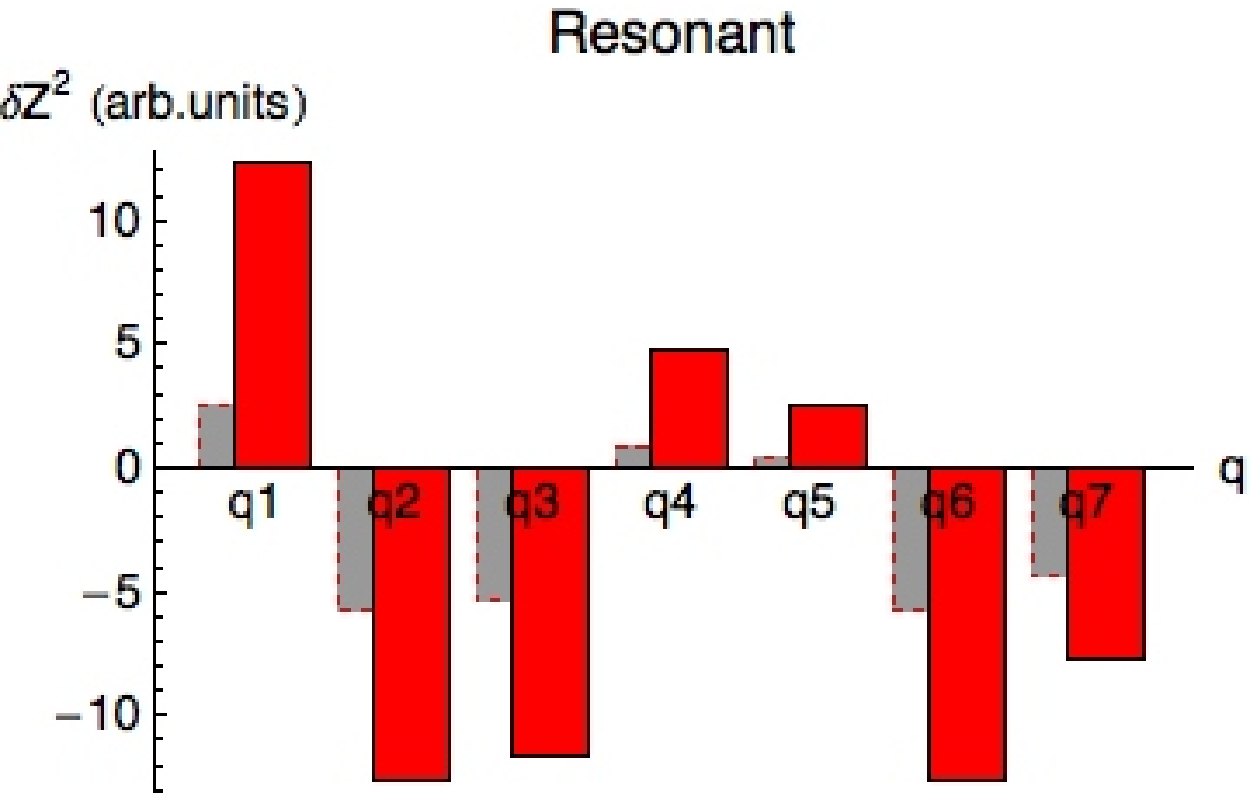}
	  }
     \hspace{.3in}
	  \subfigure[]
     {
       \label{Experiment}
		\includegraphics[width=0.45\linewidth]{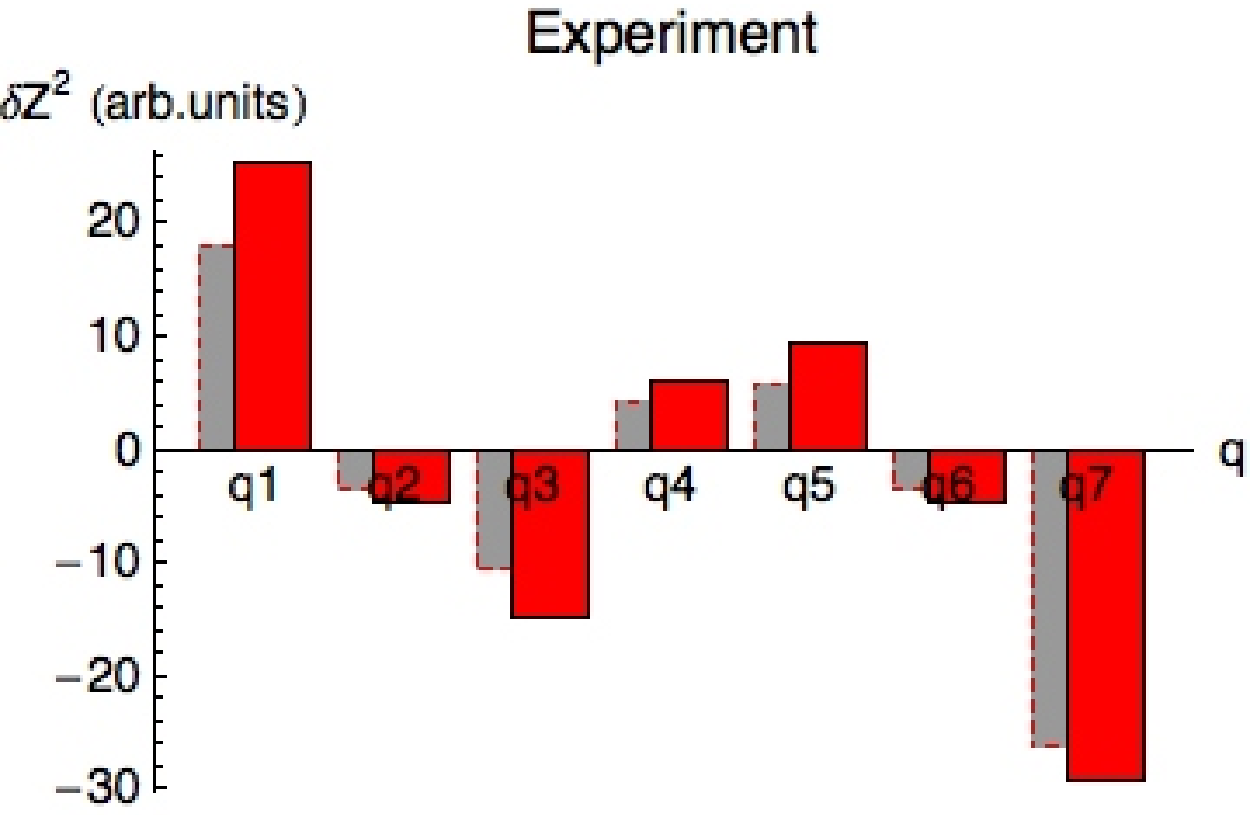}
	  }
	  \caption{(Color online) Comparison between  the results of
	  the model calculations and the experimental data.
	  Figs. (a)-(b) show the 
change in  the squared Fourier transformed conductance ratio
 $\delta Z^2\equiv|Z(\bq,V,B)|^2-|Z(\bq,V,B=0)|^2$
  at $\bq=q_{1-7}$, computed for a
magnetic field of $B=$5 T (grey bars) and 11 T (red bars)
at a bias voltage $V=\Delta_{0}/2$, provided the origin of the
selective enhancement is the Andreev (Fig. (a)) or the resonant (Fig. (b))
scattering in the vortex core region. Here a 
vortex, pinned to a scalar impurity, transforms its original
scattering potential with enhanced scattering at $q=q_{2,3,6,7}$ into an
Andreev (Fig. (a)) or into a resonant (Fig. (b)) 
scattering potential with enhanced scattering at $q=q_{1,4,5}$ (see
Table 1). 
Fig. (c) shows the experimentally observed
change in  the squared Fourier transformed conductance ratio
$\delta Z^2\equiv|Z(\bq,V,B)|^2-|Z(\bq,V,B=0)|^2$ at $\bq=q_{1-7}$, in  a
magnetic field of $B=$5 T (grey bars) and 11 T (red bars)
at a bias voltage $V=4.4$ meV.}
\end{figure}
In Fig. 3 we have modelled the scattering provided the origin of the
selective enhancement is the Andreev (Fig. (a)) or the resonant (Fig. (b))
scattering in the vortex core region. 
Both  the  Andreev and the  resonant  scattering are
equally effective in qualitatively modelling the observations.
Thus our model has qualitatively reproduced the experimentally observed
enhancement of the
sign-preserving scattering and the depression of the sign-reversing scattering.

\section{Discussion}

In this work, we have shown
how scanning tunneling spectroscopy can serve as
a phase-sensitive probe of the superconducting order parameter. In
particular, we find that 
the even and odd 
components of the density of states fluctuations can be associated 
with a well-defined coherence factor. 
The measured
Fourier transformed conductance ratio $Z(\bq,V)=\frac{dI/dV(\bq,+V)}{
dI/dV(\bq,-V)}$  is a weighted combination of these two terms, 
and in the limit of particle-hole symmetry it
is dominated by the odd component of the density of states.
Observation of coherence factor effects with scanning tunneling spectroscopy
requires the presence of controllable
scatterers. In the study by Hanaguri et al. \cite{Hanaguri} these controllable scatterers are
vortices. 

Our phenomenological model of quasiparticle scattering in the presence
of vortices is able to qualitatively reproduce the observed coherence
factor effects under the assumption that impurity scattering centers
inside the vortex cores acquire an additional Andreev or resonant
scattering component.

This study  raises several questions for
future work. 
In particular, can  a detailed model of a d-wave vortex
core provide a microscopic justification for the modification of
the impurity scattering potential? 
One of the issues that can not be resolved from the current analysis,
is whether the enhanced Andreev scattering originates in the core of
the pure vortex, ($|Z_{V}|^{2}$), or from the decoration of impurities
that are swallowed by the vortex core
($n_{core}|Z_{DI}\vert^{2}$). This is an issue that may require a combination
of more detailed experimental analysis and detailed modelling of
vortex-impurity composites using the Bogoliubov de Gennes equations. 
Another open question 
concerns whether it is  possible to 
discriminate 
between the Andreev and resonant scattering that appear to be equally
effective in accounting for the coherence factor effects.

There are several aspects to the
experimental observations that lie beyond our current work.
For example, experimentally, it is possible to spatially mask the Fourier transform data,
spatially resolving the origin of the scattering.  These masked data
provide a wealth of new information. In particular, most of the
enhancement of the sign preserving scattering is restricted to the vortex
core region, as we might expect from our theory. However, to extend
our phenomenology to encompass the masked data, requires that we
compute the fluctuations of the density of states as a function of
distance from the vortex core, 
\begin{equation}\label{}
R (\br,\br';\br_V,V)=\langle \delta \rho (\br - \br_{V},V)
\delta \rho (\br' - \br_{V},V)\rangle, 
\end{equation}
a task which requires a microscopic model of the vortex core.

In our theory we have used the bulk quasiparticle Green's functions to
compute the scattering off the vortex-decorated impurities.  Experiment
does indeed show that the quasiparticle scattering off impurities
inside the vortex cores is governed by the quasiparticle dispersion of
the bulk: can this be given a more microscopic understanding?  The
penetration of superconducting quasiparticles into the vortex core is
a feature that does not occur in conventional s-wave superconductors.
It is not clear at present to what extent  this phenomenon can be
accounted for in terms of a conservative d-wave superconductor model,
or whether it requires a more radical interpretation. One possibility
here, is that the quasiparticle fluid in both the pseudo-gap phase and
inside the vortex cores is described in terms of a ``nodal
liquid'' \cite{Balents-Fisher-Nayak}.

Beyond the cuprates, scanning tunneling spectroscopy in a
 magnetic field appears to provide a promising 
phase-sensitive probe of the symmetry of the order parameter in
unconventional superconductors. One opportunity 
that this raises, is the possibility of using STM in a field to probe
the gap phase of the newly discovered iron-based high-temperature
superconductors. 
According to one point of view \cite{Mazin}, the iron-based 
pnictide superconductors 
possess an  $s_\pm$ order parameter symmetry in which 
the order parameter has opposite signs on the hole pockets around $\Gamma$
and the  electron pockets  around M. 
If this is, indeed, the case,
then in a magnetic field quasiparticle
scattering between parts of Fermi surface with same gap signs should
exhibit  an enhancement, 
while scattering between parts of Fermi surface with opposite gap
signs will be suppressed. 
This is a point awaiting future theoretical and experimental
investigation.

We are indebted to Hide Takagi and Tetsuo Hanaguri for providing the experimental
data. We thank Hide Takagi, Tetsuo Hanaguri, J.C. Seamus Davis, Ali Yazdani,
Tami Pereg-Barnea, Marcel Franz, Peter Hirschfeld,
Zlatko Tesanovic, Eduardo Fradkin, Steven Kivelson, Jian-Xin Zhu, Sasha Balatsky
and Lev Ioffe for helpful discussions.
This research was supported by the
National Science Foundation grant  DMR-0605935.



\begin{thebibliography}{99}
\bibitem{largecores} 
Magnetization, ARPES and STM measurements consistently suggest the value of the
superconducting coherence length of about
10 times larger than the lattice constant $\xi\sim$ 10 a, far exceeding the
coherence length in BiSCCO $\xi\sim$ 2 a.
In particular, magnetization studies of optimally doped
$Ca_{2-x}Na_xCuO_2Cl_2$ \cite{Kim} 
suggest the value of upper critical field
$H_{c2}=16.9$ T which corresponds to the value of superconducting
coherence length 
$\xi_0=$44 \AA. From the average gap magnitude
$\Delta_0\sim$ 10 meV observed in the STM spectra \cite{Hanaguri,HanaguriDavis} and 
the nodal velocity of $\hbar v_F=1.8$ eV$\cdot$\AA  ~\cite{Shen, Zhou}
one obtains $\xi\sim $ 57 \AA, while 
the diameter of the vortex cores seen in STM \cite{Hanaguri}
is also of order 50 \AA $\sim$ 10 a.
\end{thebibliography}
\end{document}